\documentclass[astrosymb, fleqn, tighten]{cas-dc}

\usepackage{amsmath}
\usepackage{mathtools}
\usepackage{url}
\usepackage{gensymb}

\usepackage{cancel}
\usepackage{lineno}

\def\tsc#1{\csdef{#1}{\textsc{\lowercase{#1}}\xspace}}
\tsc{WGM}
\tsc{QE}
\tsc{EP}
\tsc{PMS}
\tsc{BEC}
\tsc{DE}
 
\newcommand{\Mearth}{${\rm M}_\oplus$}
\newcommand{\de}{{\rm d}}

\usepackage[authoryear]{natbib}

\begin{document}


\shorttitle{Circumplanetary Disks}
\shortauthors{Taylor \& Adams} 

\title[mode = title]{Formation and Structure of Circumplanetary Disks and Envelopes during the Late Stages of Giant Planet Formation} 

\author[1]{Aster~G.~Taylor}[orcid=0000-0002-0140-4475]
\cormark[1]
\fnmark[1]
\ead{agtaylor@umich.edu}
\affiliation[1]{organization={Department of Astronomy, University of Michigan},
    city={Ann Arbor},
    postcode={48109}, 
    state={MI},
    country={USA}}

\author[2,1]{Fred~C.~Adams}[orcid=0000-0002-8167-1767]
\affiliation[2]{organization={Department of Physics, University of Michigan},
    city={Ann Arbor},
    postcode={48109}, 
    state={MI},
    country={USA}}

\cortext[cor1]{Corresponding author}
\fntext[fn1]{Fannie and John Hertz Foundation Fellow}

\begin{abstract}
Giant planets are expected to form within circumstellar disks, which shape their formation history and the local environment. Here, we consider the formation and structure of circumplanetary disks that arise during the late stages of giant planet formation. During this phase, when most of the final mass is accumulated, incoming material enters the Hill sphere and falls toward the planet. In the absence of torques, the falling parcels of gas conserve their specific angular momentum and collect into a circumplanetary disk. Generalizing previous work, we consider a range of possible geometries for the flow entering the sphere of influence of the planet. Specifically, we consider five geometric patterns for the inward flow, ranging from concentration toward the rotational poles of the system to isotropic flow to concentration along the equatorial plane. For each case, we derive analytic descriptions for the density field of the infall region, the disk surface density in the absence of viscosity, and steady-state solutions for viscous disks.  These results, in turn, specify the luminosity contributions of the planet, the circumplanetary disk, and the envelope. These power sources, in conjunction with the surrounding material, collectively determine the observational appearance of the forming planet. We conclude with an approximate determination of these radiative signatures.  
\end{abstract}


\maketitle

\section{Introduction}\label{sec:intro}

Most giant planets are thought to form through the core accretion paradigm, and most of the planetary mass is accreted during the final stage of the process. Due to conservation of angular momentum, the majority of the incoming material cannot fall directly onto the planetary surface, but instead collects into a disk structure surrounding the planet. The formation, structure, and evolution of these disks thus represents a crucial part of the planet--formation paradigm. Our goal is to construct an analytic description of this process by generalizing previous work to include a wider range of boundary conditions. 

The core accretion paradigm can be broken up into three distinct stages \citep{Pollack1996}. First, rocky and icy material collects into a core with mass $M_p\simeq10\,M_\oplus$. After crossing this threshold, the protoplanet enters the second stage and gathers a hydrostatically-supported gaseous envelope extending from the core surface out to (approximately) the Hill radius. The gaseous envelope will slowly cool and collapse, lengthening the formation time while adding little mass. After the envelope mass becomes comparable to that of the rocky core, hydrostatic balance is insufficient to support the envelope. As a result, pressure no longer supports the infalling material, and the growing planet begins to gain mass in a runaway process. Although the second intermediate phase takes the longest, most of the mass of the planet is gathered during the third and final phase. As incoming material transitions from the background circumstellar disk into the sphere of influence of the planet, it retains its specific angular momentum inherited from the disk (in the absence of strong torques from magnetic fields and other mechanisms). As a result, the incoming material cannot fall directly to the planetary surface, but instead collects into a circumplanetary disk. The structure and evolution of this planet--disk system determine the properties of the forming planet during the third stage of formation. We determine the properties of infall regions and circumplanetary disks that arise during this late stage of giant planet formation.

A great deal of numerical work has been carried out concerning the problem of giant planet formation in circumstellar disks, including the third stage when rapid mass accretion occurs (e.g., \citealt{Szulagyi2016,Lambrechts2019}, and many others). These simulations often show that circumplanetary disks form alongside planets, as expected due to conservation of angular momentum (see also \citealt{Machida2008,Szulagyi2017,Fung2019}). However, even state-of-the-art numerical treatments remain somewhat limited in resolution, both temporally (with integration times much shorter than the typical $\sim10^5$ orbits of the third stage) and spatially (often using a relatively modest number of grid cells to span the Hill sphere of the forming planet), thus motivating a complementary analytic treatment.  More significantly, the results of different numerical treatments are not consistent with each other. Specifically, we are interested in the angular distribution of material entering the Hill sphere. A set of simulations finds that ``the vast majority of the mass flows into the Hill sphere near the equator'' \citep{Ayliffe2009}, while a competing study finds that ``radiative hydrodynamical simulations reveal a steady-state gas flow, which enters through the poles'' \citep{Lambrechts2017}. {Although these authors find similar meridional flow patterns, the amount of mass loaded onto a given streamline shows significant differences in these studies}. These differences {include geometrical effects and} go beyond the differing outcomes resulting from the chosen equations of state, which lead to some numerical simulations finding that disks readily form, whereas others do not. In any case, the problem of circumplanetary disk formation is not yet fully understood, {and a range geometries for the incoming flow remain possible.}

As a result, we here consider how the structure of a forming gas giant and its circumplanetary disk depend on the geometry of the inflowing material. This work expands upon an earlier treatment \citep{Adams2022} that calculated the theoretical structure of the disk and envelope for the case where flow enters the Hill sphere isotropically. 

In recent years, observational capabilities have progressed to the point that investigations of the regions immediately surrounding forming gaseous planets are feasible. Although current observations do not have the angular resolution to directly image the size scales of the planets themselves (with expected radii $R_p\sim10^{10}$ cm), they can begin to resolve circumplanetary disks (with expected sizes of $R_d\sim2\times10^{12}$ cm $\sim0.1$ AU), although such scales are currently at (or near) the resolution limit. The first tentative detection of a circumplanetary disk around a forming planet \citep{Benisty2021} is expected to be the first of many. Additional observations in the coming years will thus elucidate the formation processes for giant planets. Corresponding models of circumplanetary structure are therefore needed. The goal of this work is to provide the first step towards the requisite theoretical understanding.  

This paper is organized as follows. Sec.~\ref{sec:circstarenv} specifies the environment for forming planets, including the outer boundary conditions, as determined by the background circumstellar disk. For each of the inward flow patterns, Sec.~\ref{sec:collapse} finds the corresponding infall solutions, including the density field, the velocity field, and the column density of the envelope. These quantities determine the structure and formation of the circumplanetary disks, which are considered in Sec.~\ref{sec:diskprops}, with a focus on their surface density distributions. We present solutions for disks that are built up from the infalling flow with no further evolution and for the case of steady-state viscous accretion. In Sec.~\ref{sec:syslum}, we derive the luminosity of the planet, disk, and envelope, making several simplifying assumptions, and develop an approximate description of their spectral energy distributions. Finally, the paper concludes in Sec.~\ref{sec:conc} with a summary of our results and a discussion of their implications. 

\begin{figure}
    \centering
    \includegraphics{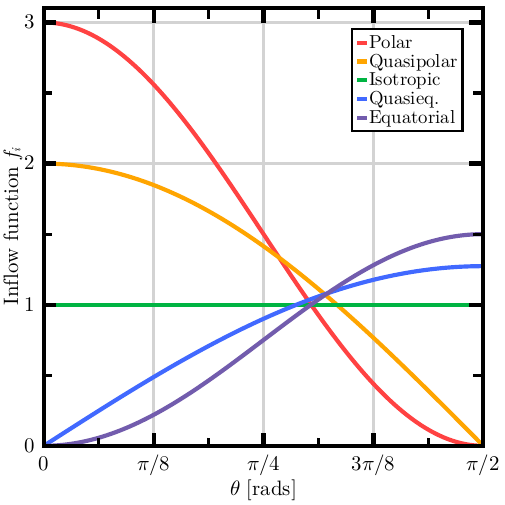}
    \vspace{-10pt}
    \caption{The infall functions used in this paper, which characterize the asymmetry in the material falling into the Hill sphere. The functions range from primarily-polar to primarily-equatorial infall. Each function is normalized such that $\oint f_i\de\Omega=1$.}
    \vspace{-10pt}
    \label{fig:inflowfuncs}
\end{figure}

\section{The Background Circumstellar Disk Environment}\label{sec:circstarenv}

In this paper, we consider the formation of gas giant planets, which actively accrete material from their parent circumstellar disks and subsequently form their own circumplanetary disks. The majority of gas giant planets are thought to form within circumstellar disks hosted by at least one star, and these disks provide the boundary conditions for the process of planet formation. As outlined above, after the growing planet reaches a mass threshold $M_p\simeq 20$ \Mearth, it can gain mass rapidly from its nearby environment. 

The boundary between the environment of the forming planet and the background circumstellar disk is nominally provided by the Hill radius $R_H=a(M_p/3M_\ast)^{1/3}$, where $a$ is the semi-major axis of the planetary orbit. However, if the Bondi radius $R_B$ = $GM/c_s^2$ is smaller than $R_H$ (where $c_s$ is the local sound speed), then $R_B$ marks the outer boundary of the planetary environment. In addition, if the scale height of the circumstellar disk $H=c_s/\Omega$ is small enough, then the outer boundary can be modified further by the thinness of the disk. In practice, the Hill radius, the Bondi radius, and the disk scale height are roughly comparable, but the Hill radius provides the boundary in most instances. In this paper, we use the Hill radius as a working approximation for the location of the boundary, but the results can easily be modified for other choices. Moreover, we will use the term ‘Hill radius’ as a proxy for this more general picture of the outer boundary. 

In order to determine how material flows toward a nascent planet, and forms a circumplanetary disk, we must specify the conditions at the outer boundary. First, we assume that the background circumstellar disk feeds material into the sphere of influence of the planet with a net rate ${\dot M}$. Numerical simulations show that some material that initially flows into the Hill sphere will subsequently flow back out. In our formulation of the problem, the mass accretion rate ${\dot M}$ only includes the material that remains within $R_H$. {Unless specified otherwise, we use a fiducial (net) mass accretion rate of 3 $M_{\rm J}$/Myr.}\footnote{{We note that larger accretion rates are sometimes advocated (e.g, \citealt{Lubow1999,DAngelo2002,Bate2003,Ayliffe2009}) and be be easily accommodated within the analytic framework developed herein.}}

To determine the size of the Hill sphere, we must specify the instantaneous mass $M_p$ of the planet along with its current orbital location $a$. As a result, the collection of variables $(a,M_p,{\dot M})$ is required to specify the boundary conditions for the planet formation problem. In addition, for a protoplanet with given mass $M_p$, we must specify its radius $R_p$ and the magnetic field strength $B_p$ on its surface, which provide inner boundary conditions.  For bodies of roughly Jovian mass, the radius is a slowly varying function of $M_p$, so we take the planetary radius to have a characteristic value $R_p=10^{10}$ cm $\simeq 1.3 R_J$ (e.g., \citealt{Marley2007}), {although young planets could in principle have radii that are even larger compared to Jupiter}. 

When its mass becomes sufficiently large, a forming planet will begin to affect the background circumstellar disk, potentially creating gaps, spiral density waves, and additional structure. The planetary mass required for gap opening depends on the viscosity of the disk and other properties (e.g., see \citealt{Fung2014}), but often occurs for masses  $M\sim50-100M_\oplus$. When planets accrete from gaps, the incoming material is no longer azimuthally symmetric, but rather enters the Hill sphere along specific directions. Because the gap region is empty, or at least has lower surface density, {the incoming material can enter the Hill sphere along directions more concentrated towards the equator, as it is accreted from the regions beyond the gap}. The advent of gap opening thus affects both the azimuthal and polar angle dependence of the incoming material. In this treatment, we retain the assumption of azimuthal symmetry (although this assumption can be relaxed using the formulation developed here). In addition to being convenient, this approximation is justified because the infalling material spirals inward to the circumplanetary disk, where it rapidly becomes evenly distributed in azimuth. Note that the orbital timescale within the circumplanetary disk is $\sim1$ yr, whereas the infall timescale is $\sim1$ Myr, so the disk has ample time to reestablish axial symmetry. On the other hand, the dependence of the incoming material on the polar angle $\theta$ can play an important role, as it determines the specific angular momentum and, hence, the location within the circumplanetary disk where the material lands.  As a result, we consider a range of $\theta$-dependent inflow patterns, as outlined below. 

Finally, we must specify the geometry of the material flowing inward through the outer boundary of the protoplanetary envelope. We consider azimuthal symmetry and latitudinal asymmetry in the inflow. More specifically, we use five different inflow functions $f_i(\theta_0)$, which set the mass flux through the Hill sphere at a polar angle $\theta_0$. These functions range from flow that is concentrated along the poles to isotropic flow to flow that is concentrated along the equatorial directions. These are written as trigonometric functions, although the absolute value is always taken, so that they are mirrored around $\theta_0=\pi$ (we will not write the absolute value signs for convenience, but always assume full symmetry). The flow functions are normalized so that the total mass inflow is $\dot{M}$, which can be enforced by setting the integral $\oint f_i d\Omega=1$. With the definition $\mu_0\equiv\cos\theta_0$, we can write the asymmetry functions for the inflow in the form 
\begin{equation}\label{eq:asyminfunc}
    f_i(\mu_0)=\begin{dcases}
        3\mu_0^2 & \text{polar},\\
        2\mu_0 & \text{quasipolar}, \\
        1 & \text{isotropic}, \\
        \frac{4}{\pi}(1-\mu_0^2)^{1/2} & \text{quasieq.}, \\
        \frac{3}{2}(1-\mu_0^2) & \text{equatorial}.
    \end{dcases}
\end{equation}
With these specifications, the mass inflow at a point on the Hill sphere is determined by $\dot{M}f_i(\mu_0)$. The asymmetry functions are shown in Fig.~\ref{fig:inflowfuncs}. 

\begin{figure}[h]
    \centering
    \includegraphics{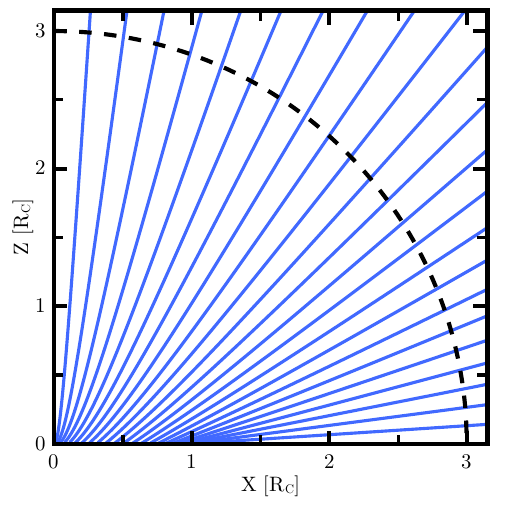}
    \vspace{-10pt}
    \caption{Infall trajectories projected into the poloidal plane, extending from the Hill sphere (dashed line) to the circumplanetary disk (in the equatorial plane along the $x$-axis). These curves are solutions to the orbit equation \eqref{eq:orbit}. The inflow is roughly spherical at large radii, but becomes more complex near $x=1\,R_C$, $z=0.1\,R_C$. The centrifugal radius $R_C$ is the outer edge of the disk.}
    \vspace{-10pt}
    \label{fig:infalltracks}
\end{figure}

\section{Infall-Collapse Solution for the Protoplanetary Envelope}
\label{sec:collapse}

With the boundary conditions and other system properties specified (see Sec.~\ref{sec:circstarenv}), we can determine how incoming material falls inward to toward the planet and its disk. 

\subsection{The Orbit Equation}

We now consider the behavior of matter once it has entered the planetary sphere of influence. As a leading-order approximation, we neglect pressure forces and assume that the infalling material travels on zero-energy ballistic orbits (see, e.g.,  \citealt{Ulrich1976,Chevalier1983,Cassen1981}). {As outlined in Appendix A of \citet{Adams2022}, under typical conditions, pressure forces correspond to a $\sim3-30\%$ correction near the Hill radius and become increasingly smaller as the material falls inward. In general, the collapse flow will tend to follow ballistic trajectories as long as cooling is efficient. In contrast, however, our solutions are not applicable in the regime of adiabatic collapse, where pressure can dominate over gravitational forces. In this case, however, the adiabatic envelope does not collapse to form a planet \citep{Fung2019}, limiting the relevance of this process.}\footnote{{See also \cite{Martin2011} for a comparison of ballistic and hydrodynamic models of disk inflow.}} 

{In the ballistic limit,} the gas must conserve its angular momentum while falling from a point on the sphere given by its initial polar angle $\theta_0$. When the gas parcel enters the Hill sphere, it has an angular velocity $\Omega$ inherited from the circumstellar disk. Its distance from the planet--disk rotation axis is given by $R_H\sin\theta_0$, and so the specific angular momentum is given by $j=\Omega R_H^2\sin^2\theta_0$. As the parcel moves inwards, it conserves its angular momentum, and thus follows an orbital equation given by 
\begin{equation}\label{eq:orbit}
    1-\frac{\mu}{\mu_0}=\zeta(1-\mu_0^2)\,.
\end{equation}
In Eq.~\eqref{eq:orbit}, the dimensionless parameter $\zeta\equiv(\Omega^2 R_H^4)/(G M_p r)$ and $\mu$ = $\cos\theta$, where $\theta$ is the polar angle in the spherical coordinate system centered on the planet.\footnote{Note that solving this equation directly for $\mu_0$ in terms of ($r$, $\mu$) requires the solution to a cubic and is thus complicated for most of the parameter space, although it is analytically exact.} For any given starting position, the parcel of gas will intersect the disk plane at $\mu=0$, shock, and join onto the disk. The size of the disk is set by the parcels with the largest specific angular momentum, which enter the Hill radius from the equatorial plane and intersect the disk at the centrifugal radius. We can solve Eq.~\eqref{eq:orbit} with $\mu=0$, $\mu_0\rightarrow0$ to find that $\zeta=R_C/r$. Assuming Keplerian orbits around the total mass $M$ that has fallen at a given epoch, we can also write $\Omega$ in terms of $G$, $M$, and $R_H$ to find that 
\begin{equation}\label{eq:RC}
    R_C=\frac{R_H}{3}\,.
\end{equation}
As the planet increases its mass, the Hill radius $R_H$ increases accordingly, and the circumplanetary disk grows so that it always extends to one third of the Hill radius \citep{Quillen1998}. The trajectories of the incoming parcels are shown in Fig.~\ref{fig:infalltracks}, where the paths are projected into the poloidal plane (see also \citealt{Tanigawa2012,Adams2022}). 

Using Eq.~\eqref{eq:orbit} and conservation of energy, one can solve for the velocity field of the collapse flow. After defining a velocity scale 
\begin{equation}\label{eq:v0}
    v_0\equiv(GM_p/r)^{1/2}\,,
\end{equation}
we can then write the velocity components in the forms
\begin{subequations}\label{eq:veqs}
    \begin{align}
        v_r=&-v_0\left[2-\zeta(1-\mu_0^2)\right]^{1/2}\label{eq:vr}\\
        v_\theta=&v_0\Big[\frac{1-\mu_0^2}{1-\mu^2}(\mu_0^2-\mu^2)\zeta\Big]^{1/2}\label{eq:vtheta}\\
        v_\phi=&v_0\Big[\frac{1-\mu_0^2}{(1-\mu^2)^{1/2}}\zeta^{1/2}\Big]\,.\label{eq:vphi}
    \end{align}    
\end{subequations}
Note that the variable $\mu_0$ is specified by Eq.~\eqref{eq:orbit} for any input location $(r, \mu)$, and so the velocity field is well defined throughout the envelope. The fact that $v_r<0$ indicates that the velocity is radially inward (see \citealt{Adams2022}; and others).

\setlength{\tabcolsep}{2pt}
\begin{table}[t]
\caption{{\textbf{Parameter Values.}} Fiducial values for model parameters, which are used as defaults throughout this paper. {While variations in these properties are possible, the analytical nature of our solutions allows the values to be changed in a straightforward manner.}}
    \centering
    \begin{tabular}{c|rl}
         \multicolumn{3}{c}{System Properties}\\\hline
         $M_\star$ & 1 & M$_\odot$ \\
         $a$ & 5 & au \\
         $\dot{M}$ & 3 & M$_{\rm J}$/Myr \\
    \end{tabular}
    \begin{tabular}{c|rl}
         \multicolumn{3}{c}{Opacity Properties}\\\hline
         $\kappa_0$ & 10 & cm$^2$ g$^{-1}$ \\
         $\nu_0$ & $10^{14}$ & Hz \\
         $\eta$ & 1 & \\
    \end{tabular}
    \begin{tabular}{c|rl}
         \multicolumn{3}{c}{Planet Properties}\\\hline
         $M_p$ & 1 & M$_{\rm J}$ \\
         $R_p$ & $10^{10}$ & cm \\
         $B_p$ & 500 & G \\
    \end{tabular}
    \label{tab:canonvals}    
\end{table}

\begin{figure*}[p]
    \centering
    \includegraphics{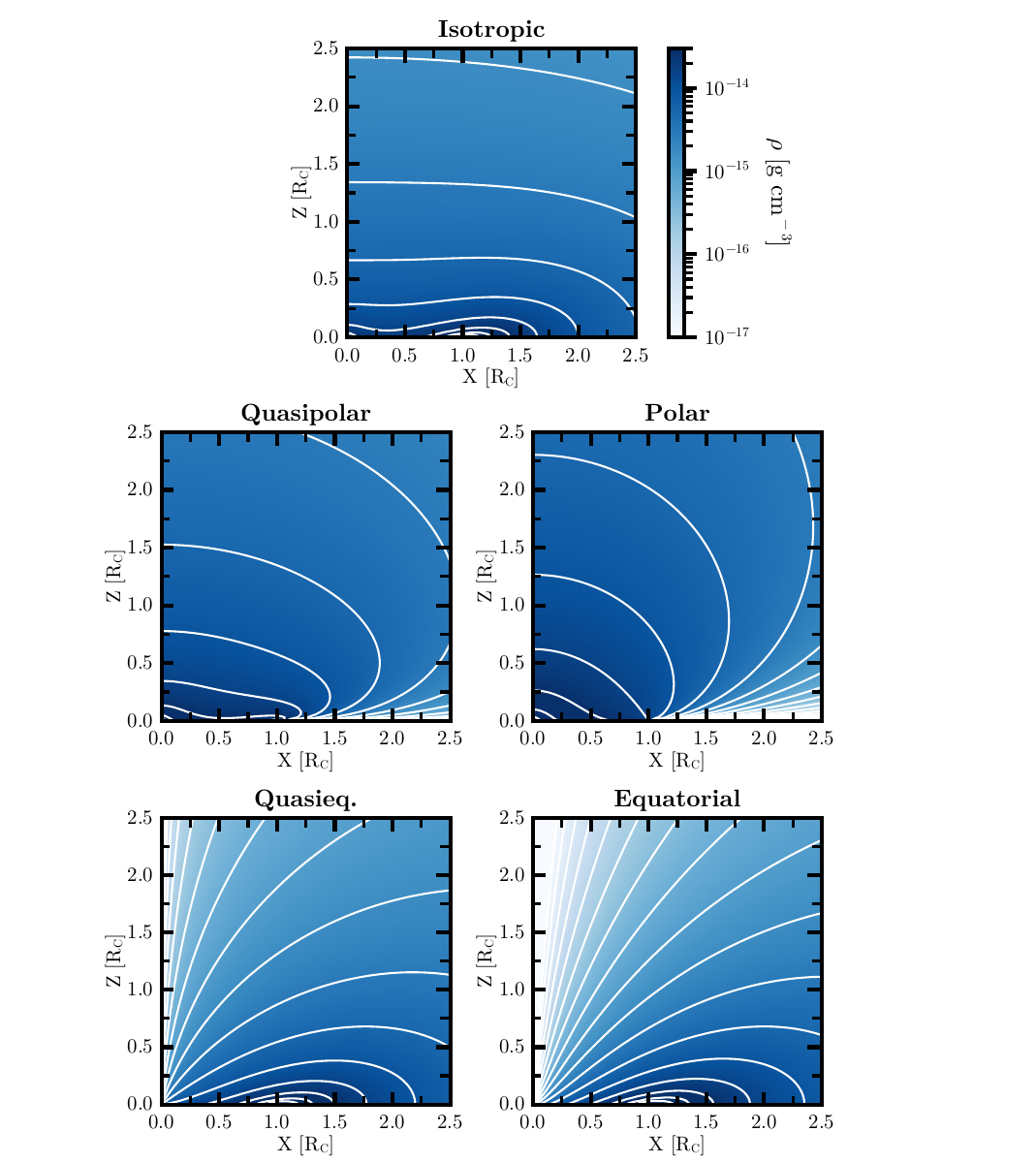}
    \vspace{-10pt}
    \caption{Density distribution of the protoplanetary envelope for each infall geometry, given by Eq.~\eqref{eq:envrhogen}. The colormap and contours are shared by all the subfigures. The contours range from $10^{-17}$ to $10^{-13}$, separated by $1/4$ in log space. {The parameter values are given by Table \ref{tab:canonvals}.}}
    \vspace{-10pt}
    \label{fig:enveloperho}
\end{figure*}

\subsection{Envelope Density Distribution}
\label{subsec:envrho}

The density of material in the envelope is given by the conservation of mass along streamlines (again see \citealt{Ulrich1976} and subsequent references). For any given point in the envelope $(r,\mu)$, we can solve Eq.~\eqref{eq:orbit} to find the initial point where the parcel crossed the Hill sphere, denoted as  $(R_H,\mu_0)$. Conservation of mass then implies that
\begin{equation}\label{eq:masscons}
\begin{aligned}    
    \rho(r,\mu)&|v_r(r,\mu)|\,\de A(r,\mu)\\=&\rho(R_H,\mu_0)|v_r(R_H,\mu_0)|\,\de A(R_H,\mu_0)\,.
\end{aligned}
\end{equation}
In Eq.~\eqref{eq:masscons}, $\de A$ is the area element at a given radius and polar angle. Since $\rho(R_H,\mu_0)=\dot{M} f_i(\mu_0)/|v_r(R_H, \mu)|$ and $\de A(r,\mu)=r^2\de\mu\de\phi$, we find that 
\begin{equation}\label{eq:envrhogen}
\begin{aligned}
    \rho(r,\mu)=&\frac{\dot{M}f_i(\mu_0)}{4\pi r^2|v_r(r,\mu)|}\frac{\de\mu_0}{\de\mu}\\
    =&\frac{\dot{M}f_i(\mu_0)}{4\pi r^2 |v_r|}[1+\zeta(3\mu_0^2-1)]^{-1}\,.
\end{aligned}
\end{equation}
In Eq.~\eqref{eq:envrhogen}, we used Eq.~\eqref{eq:orbit} to find $\de\mu_0/\de\mu$. Notice that this result is independent of the form of the inflow function $f_i$, so that extending this result using Eq.~\eqref{eq:asyminfunc} is straightforward. 

In Fig.~\ref{fig:enveloperho}, we show the resulting density distributions {for the different inflow functions, with reference parameters given in Table \ref{tab:canonvals}. Unless otherwise stated, these parameter values are used throughout the rest of this paper. The envelope density distributions} show significant differences across the inflow functions. The majority show an increased density around $r=R_H/3$, where the disk begins and a significant fraction of the infall tracks intersect the plane (see Fig.~\ref{fig:infalltracks}). However, there are significant differences in the strength of this and other qualitative features. The isotropic, quasipolar, and polar inflows show relatively similar density distributions. All three have a second point of high density near the planet, which is expected as a result of the relatively high concentration of low-angular-momentum inflow. As the inflow functions become more concentrated towards the pole, the strength of the near-planet maximum increases and the strength of the disk-edge maximum decreases, with the disk-edge maximum nearly disappearing in the fully polar case. In general, however, these density distributions are significantly more spatially uniform than in the quasiequatorial and equatorial cases.

For the flow functions that focus material toward equatorial directions, angular momentum conservation significantly concentrates the density towards the disk-edge maximum, and relatively little density remains in the direction of the planetary pole, or even the disk interior. Both of these inflow distributions then essentially lead to a toroidal circumplanetary envelope, allowing for optically thin observations of the forming planet and the disk along the poles. It is also worth noting that these density distributions require a disk to accrete material onto the planet, since almost no material falls in via the poles. Although this trend also applies in general, the effect is not as severe for polar inflow functions. 

{In this section, we have assumed azimuthal symmetry in the inflow function. This assumption is justified by the fact that the trajectories into the system depend primarily on the angular momentum, and thus only on latitude, not longitude. Due to the difference in timescales, asymmetry in the axial variable will be readily erased within the structure of both the disk and the planet. However, asymmetries in the background circumstellar disk, such as spiral density waves and/or a ``headwind'' from sub-Keplerian rotation,  could cause the envelope to only partially cover the planet \citep{Cimerman2017, Krapp2022, Kuwahara2024}. These effects are expected to be subdominant, however, and will be important mostly for the density distribution of the envelope.}

\subsection{Equivalent Spherical Infall}\label{subsec:equspherein} 

We now find the equivalent spherical infall, which is obtained by averaging the density distribution over the polar coordinate $\mu$. This average is given by 
\begin{equation}\label{eq:sphereint}
    \langle\rho\rangle=\int\displaylimits_0^1\frac{\dot{M} f_i(\mu_0)}{4\pi r^2|v_r|}\frac{\de\mu_0}{\de\mu}\,\de\mu=\int\displaylimits_\eta^1\frac{\dot{M} f_i(\mu_0)}{4\pi r^2 |v_r|}\,\de\mu_0\,.
\end{equation}
The lower limit $\eta$ of the integral depends on the radius $r$, since the streamlines that strike the disk are irrelevant. For convenience, we can define a dimensionless radius coordinate $u\equiv r/R_C$, and therefore find that
\begin{equation}\label{eq:intlim}
    \eta=\begin{dcases}
        \sqrt{1-u} & u<1;\\
        0 & u>1\,.
    \end{dcases} 
\end{equation}
The lower bound $\eta$ is defined because the streamlines begin to intersect the disk within $u<1$. As a result, there is a lower limit on the values of $\mu_0$ that contribute to the infall for $u<1$, which can be found by solving Eq.~\eqref{eq:orbit} for points where $\zeta>1$, $\mu=0$. 

The density profile that results from Eq.~\eqref{eq:sphereint} is of the general form
\begin{equation}\label{eq:sphererhoform}
    \langle\rho\rangle=Cr^{-3/2}\mathcal{A}(u)\,.
\end{equation}
In Eq.~\eqref{eq:sphererhoform}, the constant $C\equiv\dot{M}_p/4\pi\sqrt{2GM_p}$ and the asphericity function $\mathcal{A}(u)$ is defined by 
\begin{equation}\label{eq:Adef}
    \mathcal{A}(u)\equiv\sqrt{2}\int\displaylimits_\eta^1f_i(\mu_0)[2-\zeta(1-\mu_0^2)]^{-1/2}\,\de\mu_0\,.
\end{equation}
It is convenient to define 
\begin{equation}
    a^2\equiv\frac{2-\zeta}{\zeta}=2u-1\,,
\end{equation}
so that Eq.~\eqref{eq:Adef} becomes
\begin{equation}\label{eq:Asimple}
    \mathcal{A}(u)=\sqrt{2u}\int\displaylimits_\eta^1\frac{f_i(\mu_0)}{[a^2+\mu_0^2]^{1/2}}\,\de\mu_0\,.
\end{equation}

We now solve Eq.~\eqref{eq:Asimple} for the inflow functions given in Eq.~\eqref{eq:asyminfunc}. For convenience, and due to repeated patterns in the solutions, we will define a function 
\begin{equation}
    \mathcal{D}(u)=\begin{dcases}
        \ln\Big(\frac{1+\sqrt{2u}}{\sqrt{1-u}+\sqrt{u}}\Big) & u<1\,;\\
        \ln\Big(\frac{1+\sqrt{2u}}{\sqrt{2u-1}}\Big) & u>1\,.
    \end{dcases}
\end{equation}
With this definition, the isotropic case (as given in \citealt{Adams2022}; see also \citealt{Adams1986}), has the solution 
\begin{equation}
    \mathcal{A}(u)=\sqrt{2u}\mathcal{D}(u)\,.
\end{equation}

For a quasipolar inflow, we find that 
\begin{equation}
    \mathcal{A}(u)=2\sqrt{2u}\begin{dcases}
        \big[\sqrt{2u}-\sqrt{u}\big] & u<1\,;\\
        \big[\sqrt{2u}-\sqrt{2u-1}\big] & u>1.
    \end{dcases}
\end{equation}

Polar inflow produces
\begin{equation}
    \mathcal{A}(u)=\frac{3}{2}\sqrt{2u}\begin{dcases}
        \begin{aligned}
        \big[&\sqrt{2u}-(2u-1)\mathcal{D}(u)\\
        &-\sqrt{u(1-u)}\big]    
        \end{aligned} & u<1 \,;\\
        \big[\sqrt{2u}-(2u-1)\mathcal{D}(u)\big] & u>1\,.
    \end{dcases}
\end{equation}

For the quasiequatorial case, the integral must be expressed in terms of elliptic integrals. Specifically, 
\begin{equation}\label{eq:mathcalAequ}
    \mathcal{A}(u)=\frac{4}{\pi}\begin{dcases}
    \begin{aligned}
        2u\big[&F\big(\sin^{-1}(\sqrt{u}),1/2u\big)\\
        &-E\big(\sin^{-1}(\sqrt{u}),1/2u\big)\big]
    \end{aligned}  & u<1\,;\\
    \begin{aligned}
        \sqrt{2u/(2u-1)}\big[2u K(1/(1-2u)) \\
        - (2u-1)E(1/(1-2u))\big]
    \end{aligned}  & u>1\,.
    \end{dcases}
\end{equation}
In Eq.~\eqref{eq:mathcalAequ}, $K(x)$ is the complete elliptic integral of the first kind, $E(x)$ is the complete elliptic integral of the second kind, $E(x, m)$ is the incomplete elliptic integral of the second kind, and $F(x, m)$ is the incomplete elliptic integral of the first kind. Since these can be simply numerically evaluated, this form is relatively simple to use, although it is analytically complex.

\begin{figure}[t]
    \centering
    \includegraphics{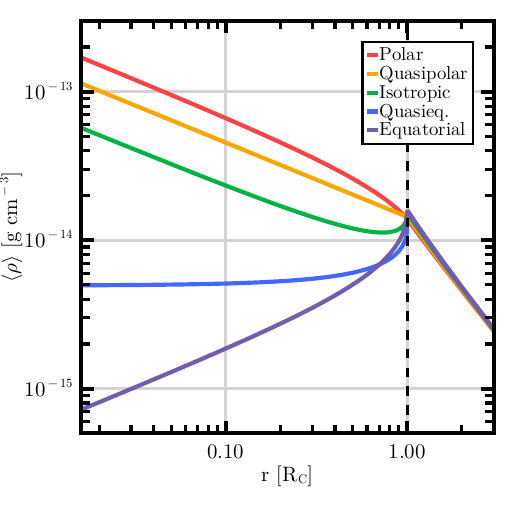}
    \vspace{-10pt}
    \caption{The equivalent spherical density for each infall function. The vertical dashed line is at $r=R_C$, the centrifugal radius/disk edge. The spike at this point is a result of a large fraction of material striking the disk and disappearing from the envelope.}
    \vspace{-10pt}
    \label{fig:sphererho}
\end{figure}

Finally, for the equatorial case, we find that 
\begin{equation}
    \mathcal{A}(u)=\frac{3}{4}\sqrt{2u}\begin{dcases}
        \begin{aligned}
        \big[&(2u+1)\mathcal{D}(u)-\sqrt{2u}\\
        &+\sqrt{u(1-u)}\big]
        \end{aligned} & u<1\,;\\
        \big[(2u+1)\mathcal{D}(u)-\sqrt{2u}\big] & u>1\,.
    \end{dcases}
\end{equation}

We show these density profiles in Fig.~\ref{fig:sphererho}. The behavior of these profiles are nearly identical beyond the disk, since the same amount of material is falling through the Hill sphere, and no trajectories have yet impacted the disk. At $r\leq R_C$, the profiles diverge significantly. The equatorial and quasiequatorial cases exhibit a significant deficit of density, as the equatorial trajectories impact the disk first (see Fig.~\ref{fig:infalltracks}). The results for the remaining flow functions show a modest feature near the centrifiugal radius, along with an shallow increase toward smaller radii. This trend is most significant for polar inflows, because relatively little mass is accreted along the equatorial streamlines which strike the disk first. 

{The density distribution has a steep slope (power-law index $p=3/2$) in the outer envelope, and a much shallower slope ($p\leq1/2$) in the inner parts $(r<R_C)$. As a result, the contribution to the column density ($N\sim\int\!\!\rho \,\de r$) is concentrated near the centrifugal radius $R_C$ for the both the outer and inner regions, and this result holds for all five geometries. Although the equatorial geometries display enhanced deficits of material in their inner regions, the net effect on the total column density is much smaller (see the following subsection), only a factor of $\sim2$, so that the corresponding effect on the spectral energy distributions is similarly modest (see Sec. \ref{sec:syslum}). }

\begin{figure}[t]
    \centering
    \includegraphics{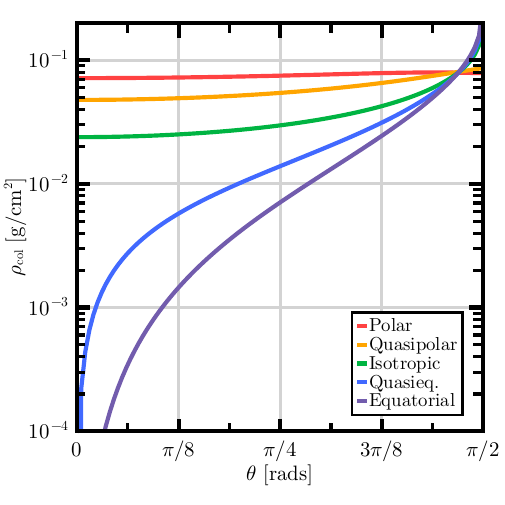}
    \vspace{-10pt}
    \caption{The column density of the envelope along a ray defined by the polar angle $\theta$. The envelope density is integrated from the planet radius $R_p$ to the Hill radius $R_H$, using the fiducial values given in Table~\ref{tab:canonvals}. Note that as $\theta\rightarrow\pi/2$, the isotropic, quasiequatorial, and equatorial infall distributions have a large column density due to concentration at the edge of the disk. }
    \vspace{-10pt}
    \label{fig:colrhonum}
\end{figure}

\begin{figure*}[t]
    \centering
    \includegraphics{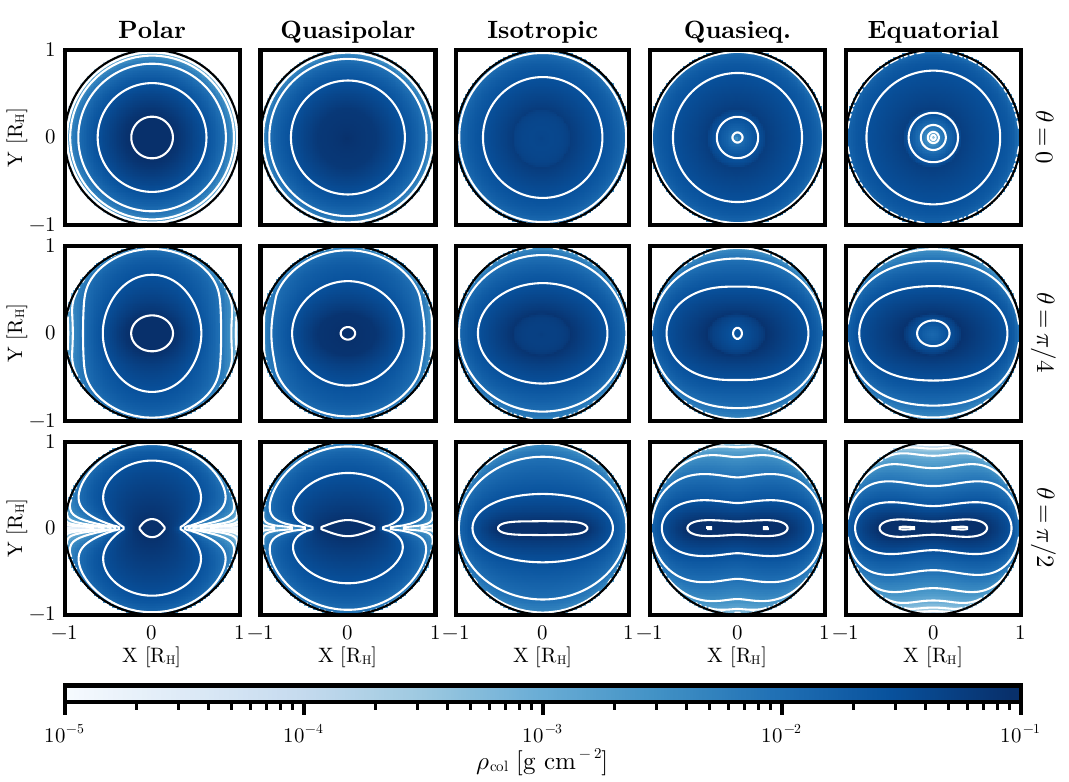}
    \vspace{-10pt}
    \caption{The column density of the envelope. The top row corresponds to a viewing angle of $0$, from above. The middle row is viewed from an angle of $\pi/4$, and the bottom row is viewed from an angle of $\pi/2$, from the equatorial plane. The color at each point represents the column density of the envelope along the ray through the system. The contours are log-spaced evenly between $10^{-4}$ and $10^{-1}$ with a step of $1/2$.  }
    \vspace{-10pt}
    \label{fig:colrhoimgs}
\end{figure*}

\subsection{Column Density}
\label{subsec:sphcolrho}

Due to the complexity of the orbit equation (Eq.~\ref{eq:orbit}), finding the column density at an arbitrary polar parameter $\mu=\cos\theta$ is not analytically tractable. However, we can numerically integrate the density (Eq.~\ref{eq:envrhogen}) along rays with different lines of sight to the center, which we present in Fig.~\ref{fig:colrhonum}. For the polar and quasipolar flow functions, the column density is relatively flat as a function of viewing angle, since the behavior of the streamlines counteracts the mass concentration. Even the isotropic case shows only modest variations with $\theta$. In contrast, the equatorial and quasiequatorial flow functions result in much more significant angular dependence, most notably a deficit of column density along the rotational poles of the system.

It is interesting to note that in Fig.~\ref{fig:colrhonum}, the polar inflow column density is essentially flat. This is a result of the fact that the infall tracks concentrate material towards the equatorial plane, while that infall distribution sets the matter to preferentially fall along the poles. These factors nearly exactly cancel. Similarly, the column densities of the other inflow functions become large at the disk plane, since material is concentrated both from the equatorial inflow and the concentration of the material. 

In order to illustrate the density structure of the system,  Fig.~\ref{fig:colrhoimgs} shows column density maps of the entire envelope along parallel lines of sight through the system. The figure shows three viewing angles --- the top row is viewed from above with $\theta=0$, the middle row is viewed from midlatitudes with $\theta=\pi/4$, and the bottom row is viewed from the equator with $\theta=\pi/2$. This figure shows the three-dimensional structure of the envelope for the different inflow distributions. In the remainder of this section, we present a simplified analytical solution, which is found by integrating through the equivalent spherical density distribution derived in Sec.~\ref{subsec:equspherein}, which provides an approximate column density for the whole system. 

Integrating through the density distribution, we evaluate 
\begin{equation}\label{eq:Ncoldef}
\begin{aligned}
    N_{\rm col}=&\int\displaylimits_0^\infty\langle\rho\rangle \de r=\int\displaylimits_0^\infty C r^{-3/2}\mathcal{A}(u)\de r\\
    =&CR_C^{-1/2}\int\displaylimits_0^\infty\frac{\sqrt{2u}}{u^{3/2}}\de u\int\displaylimits_\eta^1\frac{f(x)}{[a^2+\mu_0^2]^{1/2}}\de\mu_0\,.    
\end{aligned}
\end{equation}
For the final equality, we have used the definition of the asphericity function $\mathcal{A}(u)$. We can therefore switch the order of the integration, integrating the cosine of the angle $\mu_0$ over the range [0,1], while the integral in the radial variable $u$ develops a nonzero lower limit. We define $u_{\rm in}(x)=1-\mu_0^2$ and write Eq.~\eqref{eq:Ncoldef} as
\begin{equation}
    N_{\rm col}=C\sqrt{2R_C}\int\displaylimits_0^1f(\mu_0)\de\mu_0
    \int\displaylimits_{u_{\rm in}}^\infty\frac{\de u}{u\sqrt{2u-1+\mu_0^2}}\,.    
\end{equation}
The second integral can be evaluated and is equivalent to $(\pi/2)\,(1-\mu_0^2)^{-1/2}$. Therefore, the column density is 
\begin{equation}\label{eq:Ncolfinal}
\begin{aligned}
    N_{\rm col}=&\sqrt{2}\frac{\pi}{2}C R_C^{-1/2}\int\displaylimits_0^1\frac{f(\mu_0)\de\mu_0}{\sqrt{1-\mu_0^2}}\\
    \equiv&\sqrt{2}\frac{\pi}{2}CR_C^{-1/2}J\,.    
\end{aligned}
\end{equation}
The second equality has defined the integral $J$. Now, we simply must evaluate $J$ for the different inflow functions defined by Eq.~\eqref{eq:asyminfunc}. We find that
\begin{equation}\label{eq:J}
    J=\begin{dcases}
        3\pi/4 & \text{polar},\\
        2 & \text{quasipolar}, \\
        \pi/2 & \text{isotropic}, \\
        4/\pi & \text{quasieq.}, \\
        3\pi/8 & \text{equatorial}.
    \end{dcases}
\end{equation}

Given this result, the estimated visual extinction for a standard opacity of $\kappa_V=200$ cm$^2$ g$^{-1}$ is 
\begin{equation}
    A_V=3.3J\Big(\frac{\dot{M}}{1\,M_J/{\rm Myr}}\Big)\Big(\frac{M}{M_J}\Big)^{-2/3}\,.
\end{equation}

\section{Circumplanetary Disk Properties}
\label{sec:diskprops}

With the properties of the infalling envelope specified, we now derive the properties of the circumplanetary disk that forms at the center of the collapsing flow.

\subsection{Magnetic Truncation}
\label{subsec:magtrunc}

In general, the inner boundary of the disk will not extend all the way to the planetary surface. Instead, the disk is truncated by the planetary magnetic field. To leading order, the planetary magnetic field will have a dipole form
\begin{equation}\label{eq:planetdipole}
    \vec{B}(r,\mu)=B_p\Big(\frac{R_p}{r}\Big)^3[3\mu\hat{r}-\hat{z}]\,.
\end{equation}
The truncation radius $R_X$ is given by the radial point in the disk where the ram pressure of the inward accretion flow is balanced by the outward pressure from the magnetic field. This magnetic truncation radius has the form
\begin{equation}\label{eq:magtrunc}
    R_X=\omega\bigg(\frac{B_p^4 R_p^{12}}{GM_p\dot{M}^2}\bigg)^{1/7}\,,
\end{equation}
where $\omega$ is a dimensionless constant of order unity (for derivations of this equation, see \citealt{Ghosh1978, Blanford1982}). Inserting typical numbers, the truncation radius can be written as a scaled equation in the form 
\begin{equation}
\begin{aligned}
    \frac{R_X}{R_p}\simeq3.8&\Big(\frac{M}{M_J}\Big)^{-1/7}\Big(\frac{\dot{M}}{1\,M_J/{\rm Myr}}\Big)^{-2/7}\\
    &\times\Big(\frac{B_p}{500 \text{ Gauss}}\Big)^{4/7}\Big(\frac{R_p}{10^{10} \text{ cm}}\Big)^{5/7}\,.    
\end{aligned}
\end{equation}

\subsection{Surface Density for Direct Infall}
\label{subsec:surfacerho}

We first consider the surface density of the disk in the absence of viscous mass transfer. In this limit, the surface density at a cylindrical radius $r$ gains mass at the rate at which incoming material hits the midplane of the system. This rate is given by $2\rho v_\theta$, where the factor of $2$ arises because incoming material strikes both sides of the disk. Evaluating this quantity at $\mu=0$ using Eqs.~\eqref{eq:orbit}, \eqref{eq:veqs}, and \eqref{eq:envrhogen}, we find that 
\begin{equation}
    \frac{\de\Sigma}{\de t}=2\rho v_\theta=\frac{\dot{M}}{4\pi r^2}\frac{f_i(\mu_0)}{\zeta\mu_0}=\frac{\dot{M}}{4\pi r^2}\frac{f_i(\mu_0)}{(\zeta^2-\zeta)^{1/2}}\,.
\end{equation}
The surface density built up over time is therefore given by the integral 
\begin{equation}
    \Sigma_0(r)=\int\frac{\dot{M}}{4\pi r^2}\frac{f_i(\mu_0)}{(\zeta^2-\zeta)^{1/2}}\de t\,.
\end{equation}
Note that we can change variables by writing $\de M=\dot{M}\de t$. Since 
\begin{equation}
    \zeta=\frac{R_H}{3r}=\frac{a}{3r}\Big(\frac{M}{3M_\star}\Big)^{1/3}\,,
\end{equation}
we can also write the surface density integral in the form 
\begin{equation}\label{eq:sigma0int}
    \Sigma_0=\frac{243rM_\star}{4\pi a^3}\int\displaylimits_1^{R_C/r}\frac{f_i(\mu_0)\zeta^2}{(\zeta^2-\zeta)^{1/2}}\de\zeta\,.
\end{equation}
Given that $r<R_C$ and $\mu=0$, we can use Eq.~\eqref{eq:orbit} to write $\mu_0=\sqrt{1-1/\zeta}$.

For the isotropic case, $f_i(\mu_0)=1$, so we can evaluate the integral in Eq.~\eqref{eq:sigma0int} to find that
\begin{equation}
\begin{aligned}
    \Sigma_0=\frac{3M}{16\pi R_C^2}\Big[&\frac{(2+3u)\sqrt{1-u}}{u}\\
    &+3u\tanh^{-1}\left(\sqrt{1-u}\right)\Big]\,.    
\end{aligned}
\end{equation}
We have once again written the result in terms of $u=1/\zeta=r/R_C$.

For the quasipolar case, $f_i(\mu_0)=2\mu_0=2\sqrt{(\zeta-1)/\zeta}$. As a result, evaluating Eq.~\eqref{eq:sigma0int} yields
\begin{equation}
    \Sigma_0=\frac{3M}{4\pi R_C^2}\Big[\frac{1}{u}-u\Big]\,.
\end{equation}

For the polar case, $f_i(\mu_0)=3\mu_0^2=3(\zeta-1)/\zeta$, so that  Eq.~\eqref{eq:sigma0int} evaluates to 
\begin{equation}
\begin{aligned}
    \Sigma_0=\frac{9M}{16\pi R_C^2}\Big[&\frac{(2-u)\sqrt{1-u}}{u}\\
    &-u\tanh^{-1}\left(\sqrt{1-u}\right)\Big]\,.    
\end{aligned}
\end{equation}

For the quasiequatorial case, $f_i(\mu_0)=4\sqrt{1-\mu_0^2}/\pi=4\zeta^{-1/2}/\pi$, and we obtain 
\begin{equation}
    \Sigma_0=\frac{2M}{\pi^2 R_C^2}(1+2u)\sqrt{\frac{1-u}{u}}\,.
\end{equation}

Finally, the equatorial case gives $f_i(\mu_0)=3(1-\mu_0^2)/2=3/2\zeta$, so we evaluate Eq.~\eqref{eq:sigma0int} to find 
\begin{equation}
    \Sigma_0=\frac{9M}{8\pi R_C^2}\left(\sqrt{1-u}+u\tanh^{-1}(\sqrt{1-u})\right)\,.
\end{equation}

\begin{figure}[t]
    \centering
    \includegraphics{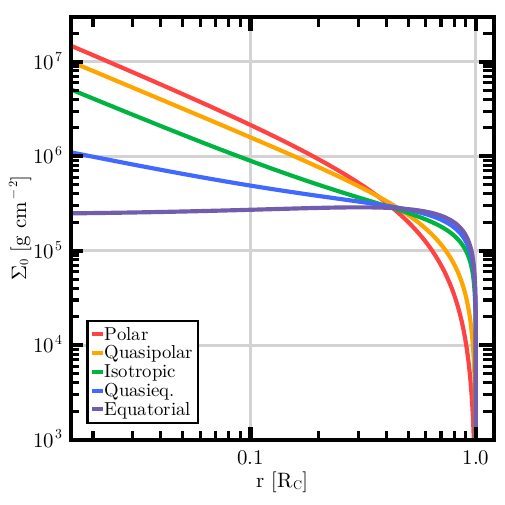}
    \vspace{-10pt}
    \caption{Surface density distributions for forming circumplanetary disks for the case of direct infall only. The curves show the results for varying input flow functions, from polar (top, red) to equatorial (bottom, purple).  All of the surface density distributions have $\Sigma_0\rightarrow0$ as $r\rightarrow R_C$, because no material impacts the equatorial plane beyond the outer edge of the disk. The inner edge is at $r=R_X$, the truncation radius. }
    \vspace{-10pt}
    \label{fig:sigmadirect}
\end{figure}

As a consistency check, we can integrate the surface density over the entire surface, that is, $\int\displaylimits_0^1 2\pi R_C^2\Sigma_0u\de u$. For all of these functions, we obtain the total mass $M$ which has come through the Hill sphere, as expected. Summarizing, we write the surface density distribution as 
\begin{equation}
    \Sigma_0=\frac{3M}{4\pi R_C^2}\begin{dcases}
        \begin{aligned}
        \frac{3}{4}\Big[&\frac{(2-u)\sqrt{1-u}}{u}\\
        -&u\tanh^{-1}\left(\sqrt{1-u}\right)\Big]
        \end{aligned}& \text{polar,} \\
        \frac{1-u^2}{u} & \text{quasipolar,} \\
        \begin{aligned}\frac{1}{4u}\big[&(2+3u)\sqrt{1-u}\\
        +&3u^2\tanh^{-1}(\sqrt{1-u})\big]\end{aligned} & \text{isotropic,} \\
        \frac{8}{3\pi}\frac{(1+2u)(1-u)^{1/2}}{\sqrt{u}} & \text{quasieq.,} \\
        \begin{aligned}
        \frac{3}{2}\big[&\sqrt{1-u}\\
        +&u\tanh^{-1}(\sqrt{1-u})\big]
        \end{aligned} & \text{equatorial.}
    \end{dcases}
\end{equation}
In the limit where $u\rightarrow0$, the isotropic and both polar solutions approach $\Sigma_0\propto u^{-1}$, the quasiequatorial solution approaches $\Sigma_0\propto u^{-1/2}$, and the equatorial solution approaches a constant, which grows even more slowly. Fig.~\ref{fig:sigmadirect} shows the behavior of these surface density functions over the radial extent of the disk.

\subsection{Steady-State Surface Density Distributions}\label{subsec:steadyrho}

Now, we include viscous accretion as a modification to the surface density. The equation of motion for a viscously-evolving disk has the form 
\begin{equation}\label{eq:diskmotionfull}
\begin{aligned}
    \frac{\partial\Sigma}{\partial t}=&\frac{3}{r}\frac{\partial}{\partial r}\bigg[r^{1/2}\frac{\partial}{\partial r}\Big(r^{1/2}\nu\Sigma\Big)\bigg]\\
    &+\frac{\dot{M}}{4\pi r^2}\frac{f_i(\mu_0)}{(\zeta^2-\zeta)^{1/2}}\,.    
\end{aligned}
\end{equation}
Eq.~\eqref{eq:diskmotionfull} includes the infall source term, and $\nu=\alpha c_s^2/\Omega$ is the kinematic viscosity. For a steady-state solution, the time derivatives vanish. Defining $S\equiv6\pi\nu\Sigma/\dot{M}$, Eq.~\eqref{eq:diskmotionfull} reduces to the steady-state form
\begin{equation}\label{eq:diskmotion}
    2\frac{\partial}{\partial u}\bigg[u^{1/2}\frac{\partial}{\partial u}\Big(u^{1/2}S\Big)\bigg]+\frac{f_i(\mu_0)}{(1-u)^{1/2}}=0\,,
\end{equation}
where $u=r/R_C=1/\zeta$ as usual. Note that the mass accretion rate through the disk is related to the function $S$ by 
\begin{equation}\label{eq:diskaccfunc}
    \dot{M}_{\rm disk}=\dot{M}u^{1/2}\frac{\partial}{\partial u}\left(u^{1/2}S\right)\,,
\end{equation}
where $\dot{M}$ is the net rate at which the planet/disk systems gains mass. For each inflow function, we can integrate Eq.~\eqref{eq:diskmotion} to find the solution for the reduced surface density $S(u)$, with integration constants $K_1$ and $K_2$, that are chosen to satisfy the boundary conditions. 

\begin{figure}[t]
    \centering
    \includegraphics{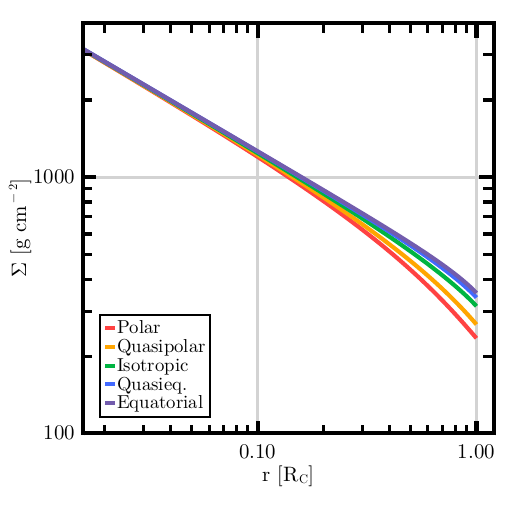}
    \vspace{-10pt}
    \caption{Steady-state surface density in a viscous disk, for the different infall functions. While the surface density stops at the edge of the disk, $r=R_C$, practically the density should continue outside of this limit. This is due to the fact that angular momentum conservation will necessarily expand the disk as viscosity operates, moving material beyond $R_C$. The inner edge is at $r=R_X$, the truncation radius.}
    \label{fig:sigmasteady}
    \vspace{-10pt}
\end{figure}

For isotropic inflow, we find that 
\begin{equation}
    S=\frac{1}{\sqrt{u}}\left[\sin^{-1}\sqrt{u}+\sqrt{u(1-u)}\right]+2K_1+\frac{K_2}{\sqrt{u}}\,.
\end{equation}
For the quasipolar inflow function, we find that 
\begin{equation}
    S=2\left[1-\frac{u}{3}\right]+2K_1+\frac{K_2}{\sqrt{u}}\,.
\end{equation}
For the polar case, the reduced density profile has the form 
\begin{equation}
\begin{aligned}
    S=\frac{1}{4}\Big[&(5-2u)\sqrt{1-u}+\frac{3}{\sqrt{u}}\sin^{-1}\sqrt{u}\Big]\\
    &+2K_1+\frac{K_2}{\sqrt{u}}\,.    
\end{aligned}
\end{equation}
For the quasiequatorial case, the profile has the form
\begin{equation}
\begin{aligned}
    S=\frac{4}{\pi}\frac{1}{\sqrt{u}}\Big[&\sqrt{u}\cos^{-1}\sqrt{u}+\frac{4}{3}-\frac{1}{3}(4-u)\sqrt{1-u}\Big]\\
    &+2K_1+\frac{K_2}{\sqrt{u}}\,.    
\end{aligned}
\end{equation}
Finally, the equatorial surface density profile takes the form of 
\begin{equation}
    S=\frac{1}{8}\left[(7+2u)\sqrt{1-u}+\frac{9}{\sqrt{u}}\sin^{-1}\sqrt{u}\right]+2K_1+\frac{K_2}{\sqrt{u}}\,.
\end{equation}
Ignoring the $K_1$ and $K_2$ terms, the steady-state surface density can be written in the form
\begin{equation}\label{eq:steadystatesurface}
    \Sigma=\frac{\dot{M}}{6\pi\nu}
    \begin{dcases}
        \begin{aligned}
           \frac{1}{4\sqrt{u}}\big[&(5-2u)\sqrt{u(1-u)}\\
            &+3\sin^{-1}\sqrt{u}\big]
        \end{aligned} & \text{polar,}\\
        2(1-u/3) & \text{quasipolar,} \\
        \frac{1}{\sqrt{u}}\big[\sqrt{u(1-u)}+\sin^{-1}\sqrt{u}\big] & \text{isotropic,} \\
        \begin{aligned}
            \frac{4}{3\pi\sqrt{u}}\big[&3\sqrt{u}\cos^{-1}\sqrt{u}\\
            &+4-(4-u)\sqrt{1-u}\big]
        \end{aligned} & \text{quasieq.,}\\
        \begin{aligned}
            \frac{1}{8\sqrt{u}}\big[&(7+2u)\sqrt{u(1-u)}\\
            &+9\sin^{-1}\sqrt{u}\big]
        \end{aligned} & \text{equatorial.}
    \end{dcases}
\end{equation}
These profiles are shown in Fig.~\ref{fig:sigmasteady}, where we assume an $\alpha$-viscosity profile \citep{Shakura1973}.

We can then evaluate Eq.~\eqref{eq:diskaccfunc} to find the limits of the accretion rate. As these limits are equivalent for all of the inflow functions, we will discuss isotropic inflow and then generalize to the other functional forms. For isotropic inflow, 
\begin{equation}
    \dot{M}_{\rm disk}=\dot{M}\big[\sqrt{1-u}+K_1\big]\,.
\end{equation}
In the limit $K_1\rightarrow0$, the disk accretion rate is thus equal to the mass inflow into the entire system (as $u\rightarrow0$). Since some fraction of the inflowing mass is accreted directly onto the planet, this provides a boundary condition for $K_1$, to ensure that the correct fraction is accreted through the disk. As $u\rightarrow1$, the mass accretion rate vanishes, since there is no material beyond $u=1$ to be brought inward. As mentioned, the other profiles behave identically at these limits with only minor differences in the leading terms. 

It is interesting to note that the surface density due to direct accretion (Fig.~\ref{fig:sigmadirect}) shows far more significant qualitative differences among the different flow functions when compared to the surface density for a steady-state viscous disk (Fig.~\ref{fig:sigmasteady}). As long as the viscosity is large enough, so that the system can reach steady-state, disk evolution smooths out the surface density into a nearly universal form. The viscous time is given by $t_{vis}\sim R_C^2/\nu$, which must be shorter than the time over which the disk mass changes from infall $t_{inf}\sim M/{\dot M}$. If we evaluate the viscous accretion time for a typical system (a planet at $a$ = 5 au in a minimum-mass solar nebula), then $t_{vis}\approx$ (16 yr)/$\alpha$, which will be much shorter than $t_{inf}\sim1$ Myr, as long as $\alpha>10^{-4}$ (see \citealt{Adams2022} for a more detailed discussion). We thus expect the circumplanetary disks to reach steady-state conditions. 

\subsection{Direct Accretion onto the Planet}\label{subsec:diracc}

In the core accretion paradigm, a certain fraction of the infalling matter is accreted directly onto the planetary surface. In this section, we calculate the fraction of material that falls onto the surface, and the planetary mass that would result in the absence of disk accretion. This mass scale is generally much smaller than the expected planetary masses, and this discrepancy underscores the importance of disk  accretion (most likely due to viscosity). 

The rate of direct infall through a sphere of radius $r<R_H$ is given by
 \begin{equation}\label{eq:Mdirectdef}
     \dot{M}_{\rm direct}=4\pi r^2\int\displaylimits_0^1\rho|v_r|\de\mu=\dot{M}\int\displaylimits_\eta^1f_i(\mu_0)\de\mu_0\,.
 \end{equation}
This result is derived using the definition of the density (Eq.~\ref{eq:envrhogen}), along with the integral limit $\eta$ given by Eq.~\eqref{eq:intlim}. 

We first consider the rate of direct infall onto the planet's surface directly, defining a fraction $f_p$ such that $\dot{M}_{\rm direct}\equiv\dot{M}f_p$, with $f_p$ given by the integral in Eq.~\eqref{eq:Mdirectdef}. By noting that $\eta=\sqrt{1-u}$ (from Eq.~\ref{eq:intlim}), we can determine that $\eta=\sqrt{1-R_p/R_C}$. Evaluating the $f_p$ integral with the inflow functions given by Eq.~\eqref{eq:asyminfunc}, we find that 
\begin{equation}\label{eq:fpvals}
    f_p=
    \begin{dcases}
        1-(1-u_p)^{3/2}&\text{polar,}\\
        u_p&\text{quasipolar,}\\
        1-\sqrt{1-u_p}&\text{isotropic,}\\
        \begin{aligned}
            &1-(2/\pi)\Big[\sqrt{u_p(1-u_p)}\\&+\arcsin{\sqrt{1-u_p}}\Big]
        \end{aligned}&\text{quasieq.,}\\
        1-\sqrt{1-u_p}(1+u_p/2)&\text{equatorial.}
    \end{dcases}
\end{equation}
Note that $u_p=R_p/R_C$. In addition, $f_d$, the fraction of material that flows directly onto the disk, is given by $f_d=1-f_p$. Working in the limit $R_C\gg R_p$, we can determine the leading order expressions for the direct infall fractions for the flow functions of interest, i.e., 
\begin{equation}\label{eq:approxints}
    f_p\simeq\begin{dcases}
        3u_p/2&\text{polar,}\\
        u_p&\text{quasipolar,}\\
        u_p/2&\text{isotropic,}\\
        (4/3\pi)u_p^{3/2}&\text{quasieq.,}\\
        (3/8)u_p^2&\text{equatorial.}
    \end{dcases}
\end{equation}

Assuming that the inflow function is consistent over the planet's lifetime, the total mass of the planet is given by 
\begin{equation}\label{eq:Mint}
    M_p=\int\displaylimits_0^tf_p\dot{M}\,\de t=M_0+\int\displaylimits_{M_0}^{M}f_p\,\de M\,.
\end{equation}
In Eq.~\eqref{eq:Mint}, $M_0$ is an initial mass value, defined to be the mass where the centrifugal radius is equal to the planetary radius, i.e., 
\begin{equation}\label{eq:M0def}
    M_0=81 M_\star\left(\frac{R_p}{a}\right)^3\,.
\end{equation}
However, the runaway gas accretion phase of giant planet formation is generally understood to begin when the planetary core mass is $M_p\simeq20$ \Mearth. Since the disk-forming mass given by Eq.~\eqref{eq:M0def} may be smaller than $20$ \Mearth, caution must be taken in determining the initial mass. However, in either case we will assume that the final mass is large compared to the initial mass, and so this correction is small.

Regardless, we will use the fact that $R_C\gg R_p$ to replace the $f_p$ in Eq.~\eqref{eq:Mint} with the approximate expressions in Eq.~\eqref{eq:approxints}. We can then use Eq.~\eqref{eq:M0def} and the fact that $R_C=(a^3 M/81M_\star)^{1/3}$ to show that 
\begin{equation}
    \frac{R_p}{R_C}=\left(\frac{M_0}{M}\right)^{1/3}\,.
\end{equation}
We use this relation to replace the $R_p/R_C$ in Eq.~\eqref{eq:approxints}. Integrating and assuming that $M_0\ll M_p$, we find that
\begin{equation}\label{eq:Mpdirsimple}
    M_p\simeq
    \begin{dcases}
        \frac{9}{4}(M_0M^2)^{1/3}&\text{polar,}\\
        \frac{3}{2}(M_0M^2)^{1/3}&\text{quasipolar,}\\
        \frac{3}{4}(M_0M^2)^{1/3}&\text{isotropic,}\\
        \frac{8}{3\pi}(M_0M)^{1/2}&\text{quasieq.,}\\
        \frac{9}{8}(M_0^2M)^{1/3}&\text{equatorial.}
    \end{dcases}
\end{equation}
Keep in mind that $M$ is the total mass of the planet--disk system, whereas $M_p$ is the mass of the planet. It is important to note, however, that this accounts only for accretion onto the planet's surface directly, and does not consider the additional effect of magnetic fields, which will expand the region where infalling matter is funneled directly onto the planet \citep{Adams2022}. Incorporation of the magnetic fields, which becomes complicated, is left for future work. 

\section{Radiation Signatures}\label{sec:syslum}

In this section, we find the luminosity of the various components of the planet--disk--envelope system. As outlined below, determining the total system luminosity is complicated by the somewhat-large number of different contributions. For the expected case where the disk viscosity is sufficiently large, however, most of the incoming material falls onto the disk surface but eventually accretes onto the central planet. The luminosity is thus dominated by the viscous contribution, along with any intrinsic luminosity within the planet itself.  With the luminosity specified, we can estimate the radiative appearance of these forming planets. Here we present an approximate treatment, but leave the full two-dimensional radiative transfer problem for a forthcoming paper. 

\subsection{Planet Luminosity}\label{subsec:planetlum}

First, let us consider the luminosity of the planet. We will assume that the planet is in thermal equilibrium, with a well-defined outer radius. As a result, the planet reemits all the power due to accretion that is delivered to its surface (where we neglect the planetary winds, which could carry some of the power). The benchmark power delivered to the entire system is the energy generated by all the material falling onto the planet's surface and dissipating its kinetic energy as if it fell from infinity. The energy scale is determined by the gravitational potential of the planet, so that the benchmark power is given by 
\begin{equation}\label{eq:L0def}
    L_0\equiv\frac{GM\dot{M}}{R_p}\,.
\end{equation}
A certain fraction of the incoming material $f_p$ falls onto the planet, while a fraction $f_d$ falls onto the disk. The planet--falling fraction $f_p$ is given by Eq.~\eqref{eq:fpvals}, while the disk-falling fraction is $f_d=1-f_p$. Note that aside from the values of the fractions $(f_p,f_d)$, the following results are general for all five inflow functions. 

The planet is assumed to be corotating with the disk at the magnetic truncation radius $R_X$, where the Keplerian rotation rate is given by 
\begin{equation}
    \Omega_p^2=\frac{GM_p}{R_X^3}\,.
\end{equation}
As a result, the matter on the planetary surface has an azimuthal rotation velocity given by 
\begin{equation}\label{eq:vphiplanet}
    v_\phi=R_p\Omega_p\sin\theta\,.
\end{equation}
In Eq.~\eqref{eq:vphiplanet}, $\theta$ is the colatitude on the planetary surface. For a given mass-inflow rate, the rate of change of kinetic energy stored in rotation has the form 
\begin{equation}
    \frac{\de E_{\rm rot}}{\de t} =\frac{1}{2}\frac{R_p^3}{R_X^3}\langle\sin^2\theta\rangle L_0\,.
\end{equation}
Assuming that the infalling material is well-mixed on the planet's surface, the average of the geometric factor $\langle\sin^2\theta\rangle=2/3$, so that the luminosity produced by the planet as a result of direct infall is
\begin{equation}\label{eq:Lpdir}
    L_p^{\rm dir}=f_p\Big(1-\frac{R_p^3}{3R_X^3}\Big)L_0\,.
\end{equation}
The luminosity produced via accretion from the disk onto the planet is similar, with a small correction. Specifically, matter no longer falls from infinity, but from the magnetic truncation radius $R_X$. As a result, the planet luminosity from material accreted from the disk is given by 
\begin{equation}\label{eq:Lpacc}
    L_p^{\rm acc}=f_d\Big(1-\frac{R_p^3}{3R_X^3}\Big)\Big(1-\frac{R_p}{R_X}\Big)L_0\,. 
\end{equation}
Adding Eqs.~\eqref{eq:Lpdir} and \eqref{eq:Lpacc}, we find that the total luminosity of the planet is 
\begin{equation}\label{eq:Lp}
    L_p=L_0\Big(1-\frac{R_p^3}{3R_X^3}\Big)\Big(1-f_d\frac{R_p}{R_X}\Big)+L_{\rm int}\,,
\end{equation}
where $L_{\rm int}$ is the intrinsic luminosity of the planet, including both power from gravitational contraction and radiogenic heating in the core. The resulting planetary temperature is given by 
\begin{equation}\label{eq:Tpdef}
    \sigma T_p^4=\frac{L_0}{4\pi R_p^2}\Big(1-\frac{R_p^3}{3R_X^3}\Big)\Big(1-f_d\frac{R_p}{R_X}\Big)+\frac{L_{\rm int}}{4\pi R_p^2}\,.
\end{equation}
In general (and specifically under the approximations of this paper), the planet's internal luminosity is small compared to the other two terms, and so that $L_{\rm int}$ can be dismissed from Eqs.~\eqref{eq:Lp} and \eqref{eq:Tpdef}. 

Assuming that the planet is a blackbody, the wavelength-dependent planet luminosity is given by 
\begin{equation}
    L_{p,\nu}=4\pi R_p^2\,\pi B_\nu(T_p)\,,
\end{equation}
and the corresponding monochromatic flux takes the form 
\begin{equation}\label{eq:Fpnu}
    F_{p,\nu}(r,\mu)=\frac{L_{p,\nu}}{4\pi r^2}G_{pd}(\mu)\exp[-\tau_\nu(r,\mu)]\,.
\end{equation}
In Eq.~\eqref{eq:Fpnu}, $\tau_\nu$ is the optical depth of the envelope between the point $(r,\mu)$ and the planet, which depends on the opacity and the column density. The function $G_{pd}(\mu)$ accounts for the possible shadowing of the disk by the planet, but is set to unity in the present treatment. Note that while the column density of the envelope is exactly specified by the solutions of the previous section, the density structure is complicated and the column density must be evaluated numerically. 

\subsection{Disk Luminosity}\label{subsec:disklum}

Next, we calculate the disk luminosity. The viscosity is expected to be large enough that most of the material landing on the disk surface will eventually accrete onto the planet (see also \citealt{Zhu2015,Zhu2018}). As a result, the total disk luminosity is $L_d=f_d L_0$, but some fraction of this power is already accounted for by the material that falls from the magnetic truncation radius to the planetary surface (specifically $L_p^{\rm acc}$). In addition, half of the energy remaining in the disk is stored in rotation due to the virial theorem, so the total disk luminosity is given by 
\begin{equation}\label{eq:Ldtot}
    L_d=f_dL_0\frac{R_p}{2R_X}\,.
\end{equation}
With the disk luminosity specified, we need to estimate the surface temperature of the disk. For both a steady-state accretion disk and a flat, passively heated disk, the temperature distribution is a power-law of the form 
\begin{equation}\label{eq:Td}
    T_d(r)=T_X\Big(\frac{R_X}{r}\Big)^{3/4}\,.
\end{equation}
Since the total luminosity of the disk is $L_d$, conservation of energy implies that 
\begin{equation}\label{eq:TXint}
    L_d=2\pi\int\displaylimits_{R_X}^{R_C}\sigma T_d^4(r)r\,\de r\,.
\end{equation}
The integral in Eq.~\eqref{eq:TXint} can be solved with Eqs.~\eqref{eq:Ldtot} and \eqref{eq:Td} to find that
\begin{equation}
    \sigma T_X^4=f_d\frac{GM\dot{M}}{8\pi R_X^3}\Big(1-\frac{R_X}{R_C}\Big)^{-1}\,.
\end{equation}

Note that we have implicitly assumed that the majority of the disk luminosity is produced by viscosity, rather than by the flow of material directly onto the disk. This approximation holds because the energy released by viscosity is the gravitational potential energy of falling from the disk onto the planet. For comparison, the energy released by the material striking the disk is the gravitational potential energy that falls onto the disk from infinity. The disk radius is much larger than the planet radius, so that the energy released by viscosity dominates over the energy released by the direct mass inflow. Moreover, the luminosity due to shocks on the disk surface and mixing of the newly added material has been calculated previously (\citealt{Adams2022}; see also \citealt{Cassen1981,Adams1986}). 

The total wavelength-dependent luminosity of the disk is given by 
\begin{equation}\label{eq:Ldnu} L_{d,\nu}=2(2\cos\theta)\int\displaylimits_{R_X}^{R_C}\pi B_\nu[T_d(r)]2\pi r\,\de r\,.
\end{equation}
We assume that the disk is optically thick to its internal radiation. The factor of $2$ arises from the two sides of the disk, and the factor of $(2\cos\theta)$ arises from a reduction of the effective area due to the viewing angle $\theta$, which must be normalized. The corresponding monochromatic flux takes the form 
\begin{equation}\label{eq:Fdnu}
    F_{d,\nu}(r,\mu)=\frac{1}{2}\frac{L_{d,\nu}}{4\pi r^2}G_{dp}(\mu)\exp(\tau_\nu(r,\mu))\,,
\end{equation}
where $G_{dp}(\mu)$ accounts for the shadowing of the disk by the planet, and $\tau_\nu$ is the optical depth of the envelope between the observation point and the disk. Note that there is a factor of $1/2$, which accounts for the fact that we assume that the disk is optically thick to itself, and so always shadows one of its faces. In addition, the use of a simple $\tau_\nu$ term for the optical depth represents an approximation --- because the disk is physically extended, the optical depth should be computed between every point on the disk surface and the observer, with the radiation from each surface element reduced by an appropriate factor. This integral is well-defined by our density solutions but is complicated to solve analytically, and is left for numerical evaluation in our forthcoming paper.

\subsection{Envelope Emission}
\label{subsec:envlum}

With the luminosity contributions specified, we estimate the spectral energy distribution for the forming planet, including the infalling envelope, the disk, and the planet itself. The envelope has no intrinsic luminosity, but it intercepts radiation from the central planet and disk and reradiates the energy at longer wavelengths. The envelope is (mostly) optically thick to the infrared radiation emitted by the planet and disk. In contrast, the envelope temperatures are lower and the reemitted radiation has longer wavelengths, so the envelope is generally optically thin to its own radiation. Since the envelope is rotating and not spherically symmetric, the resulting radiative transfer problem is two-dimensional. Leaving a full exploration of this problem for future work, we here present an approximate treatment. 

Toward this end, we use the equivalent spherical envelope approximation from Sec.~\ref{subsec:equspherein}. Since the envelope is optically thin to its own radiation, the most important property is the total column density (or optical depth) and not the specific geometry. Due to conservation of energy, the total effective envelope luminosity $L_e$ is determined by the fraction of the central source radiation that is attenuated by the infalling envelope, i.e., 
\begin{equation}\label{eq:Ledef2}
    L_e=L_p+L_d-\int\displaylimits_0^\infty (L_{p,\nu}+L_{d,\nu})\exp(-\kappa_\nu N_{\rm col})\,\de\nu\,.
\end{equation}
Note that in Eq.~\eqref{eq:Ledef2}, we have not only assumed spherical symmetry of the envelope, but placed the disk at the center of the system. In reality, of course, the disk is physically extended across the equatorial plane to $1/3$ of the envelope radius, but most of the radiation is emitted from the inner regions of the disk. In addition, the radiation emitted from the outer disk regions has a longer wavelength and is less susceptible to being absorbed. 

The effective luminosity $L_e$ of the envelope is reemitted at longer wavelengths. The total luminosity is given by the integral over the envelope volume and takes the form 
\begin{equation}\label{eq:Leint}    L_e=16\pi\int\displaylimits_{R_p}^{R_H}\rho\kappa_P\sigma T_e^4 r^2\,\de r\,,
\end{equation}
where $\kappa_P$ is the Planck mean opacity. One factor of $4\pi$ comes from the surface area term, and another comes from integration over solid angle (partially cancelled by the definition of the Stefan-Boltzmann constant $\sigma$). Because the temperature range of the envelope is limited (ranging from $T_p\sim1000$ K at the planetary surface to $T_d\sim100$ K at the Hill radius), we can assume a simple power law for the opacity,
\begin{equation}
    \kappa_\nu=\kappa_0\Big(\frac{\nu}{\nu_0}\Big)^\eta\,.
\end{equation}
The index is expected to fall in the range $1\le\eta\le2$ (see \citealt{Draine1984} and subsequent references). With this form of the frequency dependent opacity, the Planck mean opacity can be analytically evaluated and takes the form 
\begin{equation}\label{eq:kappaP}
    \kappa_P\equiv b_\kappa T^\eta=\kappa_0\Big(\frac{k_B}{\nu_0 h}\Big)^\eta\frac{\Gamma(4+\eta)\zeta_R(4+\eta)}{6\zeta_R(4)}\,T^\eta\,,
\end{equation}
where $\zeta_R$ is the Riemann zeta function and $\Gamma$ is the gamma function \citep{AbramStegun1972}. Solving the radiative transfer equations with spherical symmetry \citep{Adams1985}, and adopting an index $\eta=1$ (see also \citealt{Semenov2003}), the temperature distribution in the envelope takes the form 
\begin{equation}
    T_e(r) = T_C \left(\frac{r}{R_C}\right)^{-2/5}\,.
\end{equation}
With this temperature profile and the power-law form for the opacity, the luminosity integral (Eq.~\ref{eq:Leint}) can be evaluated to yield  
\begin{equation}\label{eq:Ledef}
L_e=16\pi R_C^2\sigma b_\kappa T_C^5 
N_{\rm col}\,,
\end{equation}
where $N_{\rm col}$ is the total column density in the spherically-equivalent limit, as defined by Eq.~\eqref{eq:Ncoldef}. However, that these are not exactly equivalent, since Eq.~\eqref{eq:Ncoldef} defines the column density to be the integral from $0$ to $\infty$, while Eq.~\eqref{eq:Leint} integrates from the planet radius $R_p$ to the Hill radius $R_H$. We use the latter result here.\footnote{Note that we are calculating the SED at the Hill radius, although additional material in the background circumstellar disk, beyond $R_H$, can process the radiation further. This correction is situationally dependent, but can be included in future applications.}

Using the amended specification of the column density $N_{\rm col}$, we can solve Eq.~\eqref{eq:Ledef} for the fiducial temperature 
$T_C$ to obtain 
\begin{equation}\label{eq:TC}
    T_C=\bigg(\frac{L_e}{16\pi R_C^2\sigma b_\kappa N_{\rm col}}\bigg)^{1/5}\,.
\end{equation}
With the fiducial temperature determined, we now have the complete (spherical) temperature profile. The total monochromatic luminosity for the envelope is then given by
\begin{equation}\label{eq:Lenu}    L_{e,\nu}=4\pi\int\displaylimits_{R_X}^{R_H}\langle\rho\rangle(r)\kappa_\nu4\pi B_\nu[T_e(r)]r^2\,\de r\,.
\end{equation}
where $\langle\rho\rangle$ is the density field of the equivalent spherical infall region, given by Sec.~\ref{subsec:equspherein}. The monochromatic flux is given by 
\begin{equation}\label{eq:Fenu}
    F_{e,\nu}=\frac{L_{e,\nu}}{4\pi r^2}\,.
\end{equation}
Note that there is no attenuation term, since the envelope is assumed to be optically thin to its own radiation. 

\subsection{Spectral Energy Distributions}\label{subsec:simpleSEDs}

Using the above results, we can now estimate the spectral energy distributions (SEDs) for forming planets. Keep in mind that we have assumed that (i) the planet and the disk are both at the center of the envelope and (ii) the envelope is spherically symmetric, using the concept of an equivalent spherical envelope. 

\begin{figure*}[t!]
    \centering
    \includegraphics{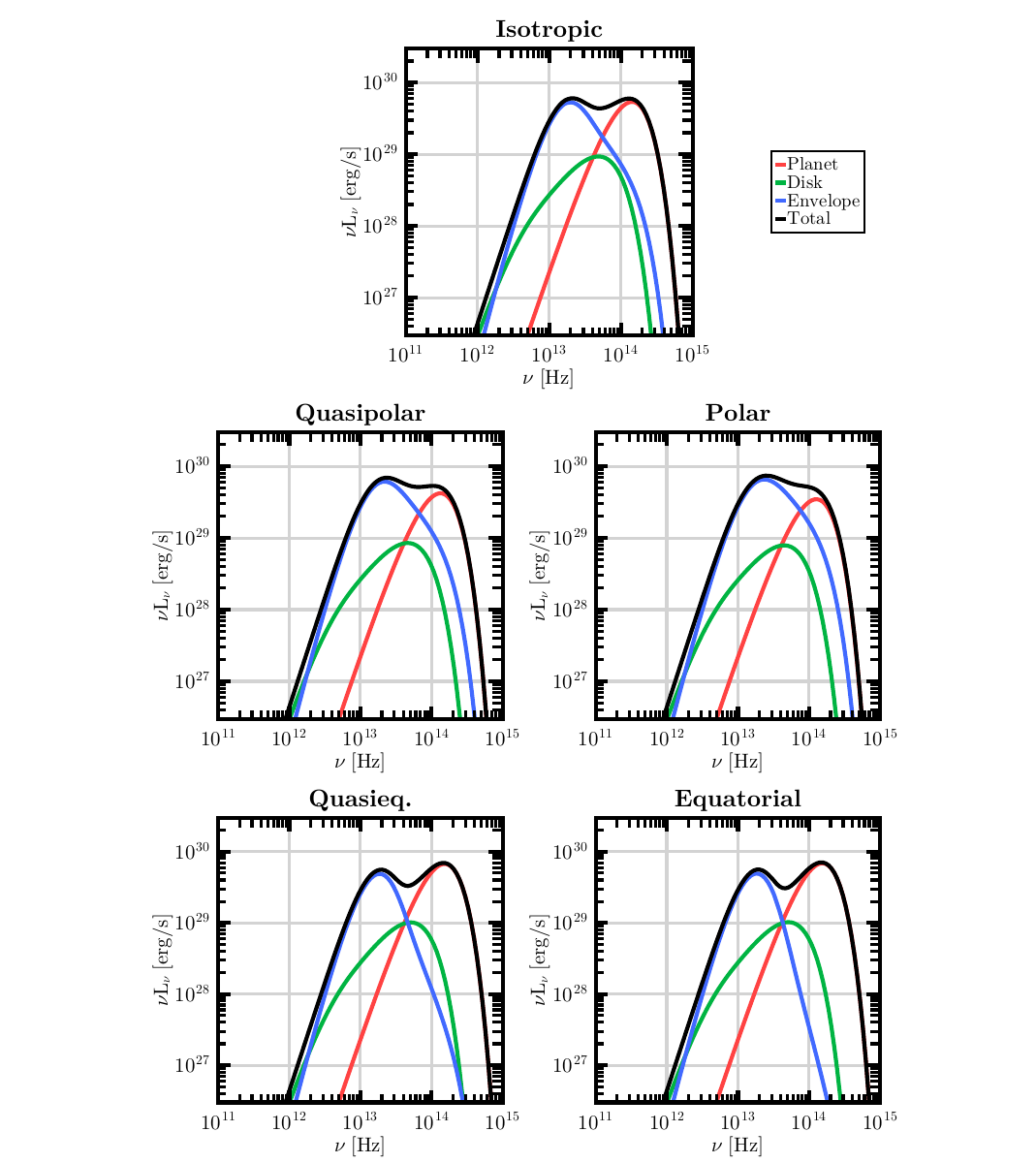}
    \vspace{-10pt}
    \caption{The spectral energy distributions of the planet--disk--envelope system, operating under the assumptions that (i) the envelope is spherically symmetric, (ii) the disk is a point source at the center of the envelope, and (iii) the envelope is optically thin to its own radiation. SEDs are shown for all 5 inflow functions. The planet and system parameter values are given in Table~\ref{tab:canonvals}. The planet, disk, envelope, and total SEDs are all shown. The viewing angle is assumed to be normal to the disk surface.   }
    \label{fig:simpleSEDinflow}
    \vspace{-10pt}
\end{figure*}

The SED is defined to be $\nu L_\nu$, where $\nu$ is the frequency. The SEDs for the representative values shown in Table~\ref{tab:canonvals} are shown in Fig.~\ref{fig:simpleSEDinflow}. It is worth noting that the disk luminosity (Eq.~\ref{eq:Ldnu}) depends on the viewing angle $\theta$. For the purposes of this figure, we assume that $\theta=0$, so that the system is seen from the rotational pole. Variation in the viewing angle is generally an $\mathcal{O}(1)$ effect on the disk SED, only becoming significant in the limit $\theta\rightarrow\pi/2$. The viewing angle has no effect on the planet or envelope contribution to the SEDs in the spherically-symmetric approximation used here. 

The spectral energy distributions presented in Fig.~\ref{fig:simpleSEDinflow} show several interesting trends. The most significant differences appear in the envelope, where the distinctions in the envelope density distribution (shown in Fig.~\ref{fig:enveloperho}) dominate. For the parameter values used here, the disk emission contribution is relatively small compared to that of the planet and the envelope. {This finding is }in apparent contrast to \cite{Szulagyi2019}, {who found that the disk is brighter than the planet at near infrared wavelengths. This difference is likely the result different choices of the inner boundary condition, especially the magnetic truncation radius, which determines the amount of energy dissipated as incoming material falls from the inner disk edge onto the planetary surface. } 

The luminosity of the planet and the disk are primarily set by the delivered accretion power, which depends on the mass inflow rate $\dot{M}$ (which is the same for all cases shown) and the fraction of matter delivered to the planetary surface $f_p$ or equivalently, that delivered to the disk $f_d=1-f_p$ (see Eq.~\ref{eq:Lp}). For the different inflow geometries, differences in the fraction $f_d$ are order $\sim u_p\sim10^{-2}$, so that this luminosity does not vary greatly. On the other hand, the differences in column density are of order unity (see Eq.~\ref{eq:J}). For these parameter values, the total optical depth of the envelope and the corresponding envelope luminosity are modestly large, so that the envelope luminosity is comparable to the unattenuated luminosity of the planet. As a result, the relative contributions of the planet and envelope vary with the infall geometry, as shown in the Fig.~\ref{fig:simpleSEDinflow}. The envelope contribution is more significant for the polar cases, primarily due to increased optical depth. 

\section{Conclusion}\label{sec:conc}

In this paper, we have developed an analytical model for the late stages of planet formation according to the core accretion paradigm, expanding on the work of \citet{Adams2022}. Those authors found theoretical results for the structure and radiative signatures of the circumplanetary disk and infalling envelope associated with a forming giant planet, assuming that the incoming mass flow is distributed isotropically. This paper extends those results to include different inflow distributions at the Hill radius, which marks the outer boundary of the system. These different boundary conditions are indicated by different numerical simulations of forming planets (compare \citealt{Ayliffe2009}, \citealt{Lambrechts2017}, and \citealt{Szulagyi2017}, which show qualitative and quantitative differences in the inflow distributions). By constructing theoretical models of this later stage of planet formation, we obtain signatures of the distribution of infalling matter that may be detectable in future observations of giant protoplanets. {Importantly, our analytical construction means that the solutions are applicable over a wide range of parameter space, independent of the specific canonical values used herein.}

The results of this paper show that the varying infall geometries {(envelope geometries)} lead to significant qualitative differences in the density structure of the infalling envelope, where the differences are greatest within the centrifugal radius (Fig.~\ref{fig:colrhonum}). The equidensity contours become prolate for infall along the polar directions and oblate for infall along the equatorial directions. The corresponding total column densities through the envelope show variations of order unity (Sec.~\ref{subsec:diracc}). The disk surface density distribution resulting from direct accretion shows corresponding distinctions (Fig.~\ref{fig:sigmadirect}), with steeper profiles for infall along the poles. In the presence of sufficiently large viscosity, however, these differences are smoothed out in the steady-state solutions of the surface density functions (Fig.~\ref{fig:sigmasteady}). As a result, the steady-state disk surface density approaches a nearly universal form as long as viscous accretion can keep up with the incoming mass flow. 

This paper also presents a preliminary assessment of the radiative appearance of forming planets for the different infall geometries under consideration. We calculate the spectral energy distributions using the ansatz of an equivalent spherical envelope, and find relatively modest observational differences (see Fig.~\ref{fig:simpleSEDinflow}). In general, the column density of the envelope is larger for flow geometries concentrated along the poles of the system (rather than the equatorial regions). As a result, the polar cases show greater luminosity from the infalling envelope, and less (unattenuated) radiation from the central planet and disk. As a result, the SEDs for the equatorial cases have a double-peaked morphology, while the polar cases vary more smoothly with frequency.  

For the canonical set of planetary parameters depicted in Fig.~\ref{fig:simpleSEDinflow}, the total optical depth of the envelope is of order unity, so that differences in the density structure have a modest effect on the SEDs. For systems with smaller optical depth (e.g., from smaller infall rate ${\dot M}$), comparatively little radiation is absorbed by the envelope, so the resulting SEDs will be similar for all geometries. For systems with larger optical depth, however, the dependence on infall geometry becomes more important. More specifically, with greater optical depth, the envelope will absorb and reradiate a greater fraction of the central source luminosity and will display a larger effective envelope luminosity. This reradiated emission depends on the density structure, and the geometry will be especially important for studying resolved systems. Notice also that for sufficiently large optical depths, the envelope will no longer be optically thin to its internal radiation. As a result, a full treatment of the envelope emission requires detailed radiative transfer calculations in two spatial dimensions, and this task is left for future work. 

{The treatment of this paper makes a number of simplifying assumptions in order to obtain analytic results. For example, the trajectories of the incoming material are calculated in the pressure-free limit. The corrections due to pressure forces are expected to be relatively modest under conditions of efficient cooling (again see the analysis of \citealt{Adams2022}), but can be substantial if the effective equation of state is sufficiently stiff. In addition, we take ${\dot M}$ to be the net infall rate, thereby ignoring material that enters the Hill sphere and immediately flows back out. This extraneous material will not affect the forming planet itself, or the disk, but its presence adds an additional component to the envelope, which in turn affects the spectral appearance of the object.  Finally, we have taken the incoming material to have azimuthal symmetry. The outer boundary conditions can break this symmetry through a variety of effects acting on the background circumstellar disk, including gap opening by the planet, headwinds due to non-Keplerian rotation, and spiral density waves. These complications will generally affect the structure of the envelope more than the disk or planet, but nonetheless should be addressed in future work.}

A number of additional avenues for future research should be explored. This present work does not include line emission, {which may be especially important when} arising from the accretion shock on the planetary surface. This radiation, especially the H$\alpha$ line, provides a metric for the system luminosity, as well as a potentially powerful diagnostic of the shock conditions on the planetary surface (e.g., \citealt{Marleau2022, Marleau2023}). Another issue that must be explored further is the effects of magnetic fields. Young and forming planets are expected to have surface fields in the range 100 -- 1000 G (e.g., \citealt{Yadav2017}), large enough to affect the geometry of the accretion flow onto the planet. In particular, the fields can enhance the cross-section for directly capturing incoming material and thereby alter the luminosity estimates. The structure of the circumplanetary disk does not only affect mass flow onto the planet and the spectral appearance of the system, but can influence the formation of satellites \citep{Batmorby2020}. Finally, when the growing planet becomes large enough to clear a gap in the circumstellar disk flow into the Hill sphere can take place along particular directions (see, e.g., \citealt{Szulagyi2022}), which will provide even more complicated boundary conditions for the incoming material. 

In this paper, we have considered the forming planet with mass $M_p$ (or system mass $M$) to have a given mass infall rate ${\dot M}$, magnetic field strength $B_p$, and semi-major axis $a$. For these values, we calculated the solutions for the density and velocity fields of the infall, the column density, the disk surface density, and so on, for each inflow geometry. In general, however, the infall rate, the magnetic field strength, and even the semi-major axis can change with time or --- equivalently --- as the planet grows in mass. Since the crossing time of the Hill sphere is on the order of $\sim1$ yr, much less than the expected evolution time  of order $\sim1$ Myr, our approach using instantaneous values of the system variables is valid. We can then incorporate evolutionary scenarios into this framework by specifying how the system properties vary with mass, although the relevant functions ${\dot M}(M_p)$, $B_p(M_p)$, and $a(M_p)$ are not yet fully understood. In addition, the flow geometry can change as the planet grows. For example, for a relatively small planet the Hill sphere will be safely smaller than the scale height of the circumstellar disk, so that an isotropic or polar geometry arises. At later times, if the Hill sphere becomes larger than the scale height, polar accretion is suppressed and the flow may become equatorial. The analytical results of this paper allow for the construction of evolutionary scenarios of this nature, including their observational consequences, although the latter must be studied in greater detail. 

\section*{Acknowledgements}

{We thank the two anonymous reviewers for their helpful comments.} We thank {Andrew Bailey, Leia Barrowes, Konstantin Batygin, Leonardo Krapp, Kaitlin Kratter, Kevin Napier, Andrew Youdin, and Zhaohuan Zhu} for useful discussions and suggestions. A.G.T. acknowledges support from the Fannie and John Hertz Foundation and the University of Michigan's Rackham Merit Fellowship Program. F.C.A. is supported in part by the Leinweber Center for Theoretical Physics at the University of Michigan. 
{This paper made use of the Julia programming language \citep{Julia}, the plotting package Makie \citep{Makie}, and Wolfram Mathematica \citep{Mathematica}.}


\bibliographystyle{icarus}
\bibliography{main.bib}

\end{document}